\documentclass{ws-p8-50x6-00}
\usepackage{amssymb,latexsym, amscd}
\usepackage[all]{xy}

 \newtheorem{theorem}{Theorem}[subsection]
 \newtheorem{cor}[theorem]{Corollary}
 \newtheorem{lemma}[theorem]{Lemma}
 \newtheorem{proposition}[theorem]{Proposition}
 \newtheorem{definition}[theorem]{Definition}
 
\usepackage{latexsym}
\usepackage{amssymb}

\newcommand{\ben}{\begin{equation}}
\newcommand{\een}{\end{equation}}

\newcommand{\integer}{\ensuremath{{\mathbb Z}}}

\newcommand{\complex}{\ensuremath{{\mathbb C}}}

\newcommand{\U}[1]{\ensuremath{{\mathrm U( #1 )}}}

\newcommand{\hyper}{\ensuremath{{\mathbb H}}}
\newcommand{\Aa}{{\mathcal A}}

\newcommand{\TT}{{\mathcal T}}

\newcommand{\UU}{{\mathcal U}}

\newcommand{\FF}{{\mathcal F}}

\newcommand{\Xx}{\mathsf{X}}
\newcommand{\sr}{\mathsf{s}}
\newcommand{\tr}{\mathsf{t}}
\newcommand{\Gg}{\mathsf{G}}
\newcommand{\Hh}{\mathsf{H}}
\newcommand{\Mm}{\mathsf{M}}
\newcommand{\Ss}{\mathsf{S}}

\newcommand{\Yy}{\mathsf{Y}}
\newcommand{\Kk}{\mathsf{K}}
\newcommand{\LLL}{\mathrm{L}}
\newcommand{\Loop}{\mathsf{L}}

\newcommand{\twoarrows}{\rightrightarrows}

\newcommand{\To}{\longrightarrow}

\newcommand{\timests}{\: {}_{t}  \! \times_{s}}

\begin{document}

\title{Deligne Cohomology for Orbifolds, discrete torsion and B-fields.}

\author{Ernesto Lupercio}

\address{University of Wisconsin, 480 Lincoln Dr., Madison WI
53706, USA\\
E-mail: lupercio@math.wisc.edu}

\author{Bernardo Uribe}

 \address{University of Wisconsin, 480 Lincoln Dr., Madison WI
53706, USA\\
E-mail: uribe@math.wisc.edu}


\maketitle

\abstracts{In this paper we introduce the concept of Deligne
cohomology of an orbifold. We prove that the third Deligne
cohomology group $H^3(\Gg,\integer(3)_D^\infty)$ of a smooth \'{e}tale
groupoid classify gerbes with connection over the groupoid. We
argue that the $B$-field and the discrete torsion in type II
superstring theories are special kinds of gerbes with connection,
and finally, for each one of them, using Deligne cohomology we
construct a flat line bundle over the inertia groupoid, namely a
Ruan inner local system\cite{Ruan} in the case of an orbifold.}

\section{Introduction}

D-brane fields in type II superstring theory can be interpreted
as a global object in $K$-theory, as has been argued by
Witten\cite{Witten}. In his work he studies the proper
quantization condition for the D-brane fields. He shows that the
fields have a charge in integral $K$-theory in much the same way
in which the electromagnetic charge of a $\U{1}$-field is the
first Chern class, namely an element in integral second
cohomology.

When we incorporate the Neveu-Schwarz $B$-field with 3-form
field strength $H$ and characteristic class $[H] \in
H^3(M,\integer)$, the corresponding statement is that the
recipient of the charge is a twisted $K$-theory group
$K_{[H]}(M)$. Witten has posed the question of  generalizing this
framework to the orbifold case. We have constructed the
appropriate recipient of the charge for types IIA and IIB
superstring theories in a previous paper\cite{LupercioUribe1}. In
that paper we argue that the corresponding version of $K$-theory
is a twisting via a \emph{gerbe} of the $K$-theory of the
\emph{stack} associated to the orbifold. We have done this for a
general orbifold not only for orbifolds known as \emph{global
quotients}, that is to say orbifolds of the form $\Xx=M/G$ with $G$
a finite group. In order to do so we have used extensively the
notion of \emph{\'etale groupoid}. This will be explained in more
detail below. The appearance of gerbes is natural even in the
smooth case.

Sharpe has argued\cite{Sharpe1} that orbifold string theories can
detect the stacky nature of the orbifold. This is consistent with
the fact that the orbifold Euler characteristic\cite{DHVW} is a
property of the stack associated to the orbifold (actually of its
inertia stack\cite{LupercioUribe2}). This point of view also fits
nicely with our construction of the orbifold $K$-theory twisted
by a gerbe.

In this paper we argue that both discrete torsion and the
$B$-field can be interpreted as classes in the Deligne cohomology
group $H^3(\Xx,\integer(3)_D^\infty)$. For this we first introduce
Deligne cohomology for orbifolds (in fact we do this for
foliations). Then we show that the Deligne cohomology group
$H^3(\Xx,\integer(3)_D^\infty)$ classifies gerbes with connection
over the orbifold (we have introduced the concept of gerbe with
connection over an orbifold in a previous
paper\cite{LupercioUribe1}).

The basic idea of interpreting the $B$-field and the discrete
torsion as Deligne cohomology classes is related to the $K$-theory
twisting in the following way. We have showed that every twisting
can be interpreted as a gerbe over the orbifold, and that gerbes
are classified by the third integral cohomology of the
classifying space $H^3(B\Xx,\integer)$. In particular we showed how
to produce such a gerbe for the case of discrete torsion. In fact
in the orbifold case the cohomology class $[H]$ corresponding to
the one used by Witten in his twisting is now a class $[H]\in
H^3(B\Xx,\integer)$. But if we want to actually consider the forms
themselves up to isomorphisms, then both the $B$-field and the
discrete torsion will then be classes in the enhanced cohomology
group $H^3(\Xx,\integer(3)_D^\infty)$. 
The mathematical expression
of the exact relation between the actual $B$-field $B \in
H^3(\Xx,\integer(3)_D^\infty)$ and its class $[H]$ is given by a
forgetting map $H^3(\Xx,\integer(3)_D^\infty) \to H^3(B\Xx,\integer)$.

Dixon, Harvey, Vafa and Witten\cite{DHVW} have shown that from
the point of view of string theory the right notion of the loop
space for the global quotient orbifold $\Xx=M/G$ should include the
twisted sectors, namely $\Loop \Xx = \coprod_{(g)} \Loop_{g} \Xx /
C(g)$ where the disjoint union runs over all the conjugacy
classes $(g)$ of elements $g\in G$, and $C(g)$ is the centralizer
of $g$ in $G$. We have generalized this argument to the case of
an arbitrary orbifold\cite{LupercioUribe2} (even one that is not a
global quotient) by means of the concept of the \emph{loop
groupoid} $\Loop \Xx $. Even in the case of global quotients the
loop groupoid has the virtue that it is invariant of the
representation chosen for the orbifold. For example if $\Xx=M_1/G_1
= M_2 /G_2$ as orbifolds then the loop groupoid will be the same
for both representations. We have also shown that the loop
groupoid $\Loop \Xx$ has a circle action, and that the $S^1$
invariant loops in $\Loop \Xx$ correspond exactly with the
twisted sectors of the orbifold that we will write as $\wedge\Xx$
(we call these the \emph{inertia groupoid}\cite{LupercioUribe2}
of the orbifold $\Xx$).

   In the case of a global quotient $\Xx=M/G$ and a discrete torsion
$\alpha\in H^3(G,\integer)$ we constructed a gerbe $\LLL_\alpha$
over $\Xx$\ (corresponding to the cohomology map
$H^3(G,\integer)\to H^3(B\Xx,\integer)$,) and then from here by
using the holonomy of this gerbe  we produced a line bundle over
$\wedge\Xx$ (corresponding to the restriction of the holonomy
line bundle over $\Loop \Xx$ of the gerbe $\LLL_\alpha$ to
$\wedge\Xx$).

In this paper we prove that such line bundle admits a natural
\emph{flat} connection. In fact we prove something more
interesting and general. For \emph{any} gerbe with a connection
over $\Xx$ we will construct a \emph{flat} line bundle over the
twisted sectors $\wedge\Xx$. We will do this using Deligne
cohomology in the following way. We will prove that
$H^3(\Xx,\integer(3)_D^\infty)$ classifies \emph{gerbes with
connection} over $\Xx$, that $H^2(\Yy,\integer(2)_D^\infty)$
classifies \emph{line bundles with connection} over $\Yy$, and
finally that there exists a natural holonomy map
\begin{equation}\label{holonomy}
H^3(\Xx,\integer(3)_D^\infty) \longrightarrow
H^2(\wedge\Xx,\integer(2)_D^\infty)
\end{equation}
whose image lands in the family of \emph{flat} line bundles over
$\wedge\Xx$.

To finish this introduction let us mention that the mathematical
results of this paper can be approached from a more general point
of view that includes all the Deligne cohomology groups (not only
the second and the third) and their geometric
interpretations\cite{LupercioUribe3}.

\section{Preliminaries}
In this section we will review the basic concepts of groupoids, sheaves over groupoids
and the cohomologies associated to them. For a more detailed description we recommend
to see Haefliger\cite{Haefliger}, Crainic and Moerdijk\cite{CrainicMoerdijk} and Lupercio
and Uribe\cite{LupercioUribe2}.

\subsection{Topological Groupoids}\label{subsec:groupoids}

The groupoids we will consider are small categories $\Gg$ in which every morphism
 is invertible. By $\Gg_1$ and $\Gg_0$ we will denote the space of morphisms (arrows)
and of objects respectively, and the structure maps

        $$\xymatrix{
        \Gg_1 \timests \Gg_1 \ar[r]^{m} & \Gg_1 \ar[r]^i &
        \Gg_1 \ar@<.5ex>[r]^s \ar@<-.5ex>[r]_t & \Gg_0 \ar[r]^e & \Gg_1
        }$$
where $s$ and $t$ are the source and the target maps, $m$ is the composition of two arrows,
$i$ is the inverse and $e$ gives the identity arrow over every object.

 The groupoid will be called {\it topological (smooth)} if the sets $\Gg_1$ and $\Gg_0$
 and the structure maps belong to the category of topological spaces (smooth manifolds).
In the case of a smooth groupoid we
will also require that the maps $s$ and $t$ must be submersions, so that
$\Gg_1 \timests \Gg_1 $ is also a manifold.

A topological (smooth) groupoid is called {\it \'{e}tale} if all the structure maps
are local homeomorphisms (local diffeomorphisms). For an \'{e}tale groupoid we will
mean a topological \'{e}tale groupoid. In what follows, sometimes the kind of groupoid
will not be specified, but it will be clear from the context to which one we are referring to.
We will fix notation for the groupoids and they will be denoted only by
letters of the type $\Gg,\Hh,\Ss$.

Orbifolds are an special kind of \'{e}tale groupoids, they have the peculiarity
that the map $(s,t): \Gg_1 \to \Gg_0\times \Gg_0$ is proper, groupoids
with this property are called {\it proper}. Whenever we write
orbifold, a proper \'{e}tale smooth groupoid will be understood.

A morphism of groupoids $\Psi: \Hh \to \Gg$ is a pair of maps
$\Psi_i: \Hh_i \to \Gg_i$ $i=1,2$ such that they commute with the
structure maps. The maps $\Psi_i$ will be continuous (smooth)
depending on which category we are working on.

The morphism $\Psi$ is called {\it Morita} if the following square is a cartesian square
       $$\xymatrix{
         \Hh_1 \ar[r]^{\Psi_1} \ar[d]_{(s,t)} & \Gg_1 \ar[d]^{(s,t)} \\
         \Hh_0 \times \Hh_0 \ar[r]^{\Psi_0 \times \Psi_0} & \Gg_0 \times \Gg_0.
         }$$
If the Morita morphism is between \'{e}tale groupoids and the map
$s \pi_2:\Hh_0 \: {}_{\Psi_0}  \! \times_{t} \Gg_1 \to \Gg_0$ is
an \'{e}tale surjection (local homeomorphism, diffeomorphism)
then the morphism is called {\it \'{e}tale Morita}; for orbifolds
we will require the Morita morphisms to be \'{e}tale.

Two groupoids $\Gg$ and $\Hh$ are Morita equivalent if there
exist another groupoid $\Kk$ with Morita morphisms $\Gg
\stackrel{\simeq}{\leftarrow} \Kk \stackrel{\simeq}{\to} \Hh$.

For an \'{e}tale groupoid $\Gg$, we denote by $\Gg_n$ the space of $n$-arrows
$$x_0 \stackrel{g_1}{\to} x_1 \stackrel{g_2}{\to} \cdots \stackrel{g_n}{\to} x_n$$
The spaces $\Gg_n$ $(n \geq 0)$ form a simplicial space:

        $$\xymatrix{
        \cdots \ar@<1.5ex>[r] \ar@<.5ex>[r] \ar@<-.5ex>[r] \ar@<-1.5ex>[r] &
        \Gg_2 \ar@<1ex>[r] \ar[r] \ar@<-1ex>[r] &
        \Gg_1 \ar@<.5ex>[r]^s \ar@<-.5ex>[r]_t & \Gg_0
        }$$
that together with the face maps $d_i: \Gg_n \to \Gg_{n-1}$
$$d_i(g_1, \dots, g_n) = \left\{
\begin{array}{cc}
(g_2, \dots, g_n) & \hbox{if $i=0$}\\
(g_1, \dots, g_ig_{i+1}, \dots, g_n) & \hbox{if $0 <  i <n$}\\
(g_1, \dots ,g_{n-1}) & \hbox{if $i=n$}
\end{array} \right.$$
form what is called the {\it nerve of the groupoid}. Its geometric realization\cite{Segal1}
 is the classifying space of $\Gg$, denoted $B\Gg$. A Morita
equivalence $\Hh \stackrel{\sim}{\to} \Gg$ induces a weak homotopy equivalence
$B\Hh \stackrel{\sim}{\to} B\Gg$.

\subsection{Sheaves and cohomology}\label{subsec:sheaves}

From now on, we will restrict our attention to the case where $\Gg$ is
an \'{e}tale groupoid and smooth when required.

A $\Gg$-sheaf $\FF$ is a sheaf over $\Gg_0$, namely a topological space
with a projection $p: \FF \to \Gg_0$ which is a local homeomorphism
on which $\Gg_1$ acts continuously. This means that for $a \in \FF_x=p^{-1}(x)$
and $g \in \Gg_1$ with $s(g)=x$, there is an element $ag$ in $\FF_{t(g)}$
depending continuously on $g$ and $a$. The action is a map
$\FF \: {}_{p} \! \times_s \Gg_1 \to \FF$.

All the properties of sheaves and its cohomologies of topological spaces can be
extended for the case of \'{e}tale groupoids as is done in Haefliger\cite{Haefliger}
and Crainic and Moerdijk\cite{CrainicMoerdijk}.

For $\FF$ a $\Gg$-sheaf, a section $\sigma : \Gg_0 \to \FF$ is called invariant
if $\sigma(x) g =\sigma(y)$ for any arrow $x \stackrel{g}{\to} y$. $\Gamma_{inv}(\Gg,\FF)$
is the set of invariant sections and it will be an abelian group if $\FF$ is an abelian sheaf.

For an abelian $\Gg$ sheaf $\FF$, the {\it cohomology} groups $H^n(\Gg,\FF)$ are defined as
the cohomology groups of the complex:
$$\Gamma_{inv}(\Gg, \TT^0) \to \Gamma_{inv}(\Gg,\TT^1) \to \cdots$$
where $\FF \to \TT^0 \to \TT^1 \to \cdots$ is a resolution of $\FF$ by injective
$\Gg$-sheaves. When the abelian sheaf $\FF$ is locally constant (for example
$\FF =\integer$) is a result of Moerdijk\cite{Moerdijk} that
$$H^*(\Gg,\FF) \cong H^*(B\Gg, \FF)$$
where the left hand side is sheaf cohomology and the right hand side is simplicial
cohomology.

There is a {\it basic spectral sequence} associated to this cohomology. Pulling back
$\FF$ along
\begin{eqnarray}
\epsilon_n : \Gg_n \to \Gg_0 \label{pullbacksheaves}
\end{eqnarray}
 $$\epsilon_n(g_1, \dots, g_n) = t(g_n)$$ it
induces a sheaf $\epsilon_n^*\FF$ on $\Gg_n$ (where the $\Gg$ action on $\Gg_n$ is
the natural one, i.e. $(g_1, \dots, g_n)h= (g_1, \dots, g_nh)$; $\Gg_n$ becomes in
this  way a $\Gg$-sheaf) such that for fixed $q$ the groups $H^q(\Gg_p , \epsilon_p^* \FF)$
form a cosimplicial abelian group, inducing a spectral sequence:
$$H^pH^q(\Gg_\bullet, \FF) \Rightarrow H^{p+q}(\Gg,\FF)$$
So if $0 \to \FF \to \FF^0 \to \FF^1 \to \cdots$ is a resolution of $\Gg$-sheaves with
the property that $\epsilon_p^* \FF^q$ is an acyclic sheaf on $\Gg_p$, then $H^*(\Gg,\FF)$
can be computed from the double complex $\Gamma(\Gg_p, \epsilon_p^* \FF^q)$.

We conclude this section by introducing the algebraic gadget that will allow us to define
Deligne cohomology. Let $\FF^\bullet$ be a cochain complex of
abelian sheaves, then the {\it hypercohomology groups}
$\hyper^n(\Gg,\FF)$ are defined  as the cohomology groups of the
double complex $\Gamma_{inv}(\Gg, \TT^\bullet)$ where $\FF^\bullet \to \TT^\bullet$ is a
quasi-isomorphism into a cochain complex of injectives.

\section{Deligne Cohomology}
In what follows we will define the smooth Deligne cohomology of a smooth \'{e}tale
groupoid; we will extend the results of Brylinski\cite{Brylinski}  
to groupoids and will follow very closely the description given in there.

We will assume all through out this paper that the set of objects $\Gg_0$ of our
groupoid $\Gg$ has an open cover by subsets which are each paracompact, Hausdorff,
locally compact and of bounded cohomological dimension depending on $\Gg_0$.
If not specified explicitly, when working on a groupoid we will have always in mind
its description given by this cover; in other words, we will think of $\Gg_0$ as
the disjoint union of this cover and $\Gg_1$ its respective space of arrows that will
keep the groupoid in the same Morita class.

Deligne cohomology is related to the De Rham cohomology. We will consider the De Rham complex
of sheaves and we will truncate it at level $p$; what interests us is the degree
$p$ hypercohomology classes of this complex. To be more specific,  let $\integer(p):=
(2 \pi \sqrt{-1})^p \cdot \integer$ be the cyclic subgroup of $\complex$, $A^p(\Gg_0)_\complex:=
A^p(\Gg_0) \otimes \complex$ the space of complex-valued $p$ forms on $\Gg_0$ and
$\Aa^p_{\Gg, \complex}$ the $\Gg$-sheaf of complex-valued differential $p$-forms; as
$\Gg$ is a smooth \'{e}tale groupoid the maps $s$ and $t$ are local diffeomorphisms,
 then the action  of $\Gg$ into the sheaf over $\Gg_0$
of complex-valued differential $p$-forms is the natural one given by the pull back of
the corresponding diffeo.

Let $\integer(p)_\Gg$ be the constant $\integer(p)$-valued $\Gg$-sheaf, and
$i : \integer(p)_\Gg \to \Aa^0_{\Gg,\complex}$ the inclusion of constant
into smooth functions.

\begin{definition}
Let $\Gg$ be a smooth \'{e}tale groupoid. The {\bf smooth Deligne complex}
$\integer(p)_D^\infty$ is the complex of $\Gg$-sheaves:
$$\integer(p)_\Gg \stackrel{i}{\To} \Aa^0_{\Gg,\complex} \stackrel{d}{\To}
\Aa^1_{\Gg,\complex} \stackrel{d}{\To} \cdots \stackrel{d}{\To} \Aa^{p-1}_{\Gg,\complex}$$
The hypercohomology groups $\hyper^q(\Gg,\integer(p)_D^\infty)$ are called
the {\bf smooth Deligne cohomology} of $\Gg$.
\end{definition}

Formally, this description will do the job, but we would like to have a more concrete
definition of this cohomology. This is done using a {\it Leray} description of
the groupoid  and then calculating the cohomology of the respective
\u{C}ech double complex. For a manifold $M$, a {\it Leray} cover $\UU$ is one
on which all the open sets and its intersections are contractible. With this
cover we can calculate the De Rham cohomology of $M$ by calculating the cohomology
of the \u{C}ech complex with real coefficients; for the De Rham cohomology
of the open sets in the cover is trivial.
 The same idea can be applied to the hypercohomology of the
Deligne complex on a manifold, but in this case we obtain a \u{C}ech double complex.
This will be explained in more detail in the next section.

\begin{definition}
Let $\Hh$ be a smooth \'{e}tale groupoid. A {\it Leray} description of $\Hh$ is an
\'{e}tale Morita equivalent groupoid $\Gg$, provided with an \'{e}tale 
Morita morphism $\Gg \to \Hh$, on
which all the sets $\Gg_i$, for $i \geq 0$, are diffeomorphic to a disjoint union
of contractible sets.
\end{definition}

The existence of such a groupoid for orbifolds (namely proper \'{e}tale smooth
groupoids) can be proved using the results of Moerdijk and Pronk \cite{MoerdijkPronk}.
In any case, this is clear for most relevant examples.
In order to make the calculations clearer, where are going to work
with a quasi-isomorphic complex
of sheaves to the Deligne one, which is a bit simpler.

\begin{definition} Let $\complex^\times(p)_\Gg$ be the following complex of sheaves:
$$\complex^\times_\Gg \stackrel{d \log}{\To} \Aa^1_{\Gg,\complex} \stackrel{d}{\To}
\cdots \stackrel{d}{\To} \Aa^{p-1}_{\Gg,\complex}$$
\end{definition}

It's easy to see that there is a quasi-isomorphism between the  complexes
$(2\pi \sqrt{-1})^{-p+1}\cdot \integer(p)_D^\infty$ and $\complex^\times(p)_\Gg[-1]$
(this fact is explained in Brylinski \cite{Brylinski} page 216)

    $$\xymatrix{
    (2\pi \sqrt{-1})^{-p+1}\cdot\integer(p)_\Gg \ar[r] &
     \complex_\Gg \ar[r]^{d} \ar[d]_\exp &\Aa^1_{\Gg,\complex} \ar[r]^{d} \ar[d] &
\cdots \ar[r]^{d} & \Aa^{p-1}_{\Gg,\complex} \ar[d]\\
    &    \complex^\times_\Gg \ar[r]^{d \log} &\Aa^1_{\Gg,\complex} \ar[r]^{d} &
\cdots \ar[r]^{d} &\Aa^{p-1}_{\Gg,\complex}
    }$$
hence there is an isomorphism of hypercohomologies:
\begin{eqnarray} \label{isodelignecohomology}
\hyper^{q-1}(\Gg, \complex^\times(p)_\Gg) \cong (2\pi \sqrt{-1})^{-p+1} \cdot
\hyper^q(\Gg, \integer(p)_D^\infty)
\end{eqnarray}

Now let $\Gg$ be such Leray description of the groupoid. We are going to define the \u{C}ech
double complex associated to the $\Gg$-sheaf complex $\complex^\times(p)_\Gg$. Consider the space
$$C^{k,l}=\breve{C}(\Gg_k, \Aa^l_{\Gg,\complex}):=\Gamma(\Gg_k, \epsilon_k^*\Aa^l_{\Gg,\complex})$$
 of global sections of the sheaf $\epsilon_k^*\Aa^l_{\Gg,\complex}$ over
$\Gg_k$ as in (\ref{pullbacksheaves}).
The vertical differential $C^{k,l} \to C^{k,l+1}$ is given by the maps of the complex
$\complex^\times(p)_\Gg$
 and the horizontal differential $C^{k,l} \to C^{k+1,l}$ is
obtained by $\delta = \sum(-1)^i\delta_i$
where for $\sigma \in \Gamma(\Gg_k, \epsilon_k^*\Aa^l_{\Gg,\complex})$

$$ (\delta_i \sigma)(g_1,\dots,g_{k+1}) = \left\{
\begin{array}{cc}
\sigma(g_1, \dots, g_k)\cdot g_{k+1} & \hbox{for $i=k+1$}\\
\sigma(g_1, \dots, g_ig_{i+1}, \dots, g_{k+1}) & \hbox{for $0<i<k+1$}\\
\sigma(g_2,\dots,g_{k+1}) & \hbox{for $i=0$}
\end{array} \right.$$

\begin{definition}
For $\Gg$ a Leray description of a smooth \'{e}tale groupoid, let's
denote by $\breve{C}(\Gg,\complex^\times(p)_\Gg)$ the total complex induced by the double complex

   $$\xymatrix{
     \vdots & \vdots & \vdots & & \vdots\\
     \breve{C}(\Gg_2, \complex^\times_\Gg) \ar[u]^\delta \ar[r]^{d \log} &
         \breve{C}(\Gg_2, \Aa^1_{\Gg,\complex})
        \ar[r]^d \ar[u]^\delta &  \breve{C}(\Gg_2, \Aa^2_{\Gg,\complex})
        \ar[u]^\delta \ar[r]^d & \cdots \ar[r]^d & \breve{C}(\Gg_2, \Aa^{p-1}_{\Gg,\complex})
         \ar[u]^\delta\\
      \breve{C}(\Gg_1, \complex^\times_\Gg) \ar[u]^\delta \ar[r]^{d \log} &
          \breve{C}(\Gg_1, \Aa^1_{\Gg,\complex})
        \ar[r]^d \ar[u]^\delta &  \breve{C}(\Gg_1, \Aa^2_{\Gg,\complex})
        \ar[u]^\delta \ar[r]^d & \cdots \ar[r]^d & \breve{C}(\Gg_1, \Aa^{p-1}_{\Gg,\complex})
        \ar[u]^\delta\\
      \breve{C}(\Gg_0, \complex^\times_\Gg) \ar[u]^\delta \ar[r]^{d \log} &
         \breve{C}(\Gg_0, \Aa^1_{\Gg,\complex})
          \ar[r]^d \ar[u]^\delta &  \breve{C}(\Gg_0, \Aa^2_{\Gg,\complex})
          \ar[u]^\delta \ar[r]^d & \cdots\ar[r]^d & \breve{C}(\Gg_0, \Aa^{p-1}_{\Gg,\complex}) 
          \ar[u]^\delta
    }$$
The \u{C}ech hypercohomology of the complex of sheaves $\complex^\times(p)_\Gg$ is defined
as the cohomology of the \u{C}ech complex $\breve{C}(\Gg,\complex^\times(p)_\Gg)$:
$$\breve{H}^*(\Gg, \complex^\times(p)_\Gg):=H^*\breve{C}(\Gg,\complex^\times(p)_\Gg).$$
\end{definition}

Due to all the conditions imposed to the Leray description, the previous cohomology
actually matches the hypercohomology of the complex $\complex^\times(p)_\Gg$, so we get
\begin{lemma} \label{cech=hyper}
Let $\Hh$ be a smooth \'{e}tale groupoid and $\Gg$ a Leray description of it. Then
the cohomology of the \u{C}ech complex $\breve{C}(\Gg,\complex^\times(p)_\Gg)$ is
isomorphic to the hypercohomology of $\complex^\times(p)_\Hh$
$$\breve{H}^*(\Gg, \complex^\times(p)\Gg) \stackrel{\cong}{\to}
\hyper^*(\Gg,\complex^\times(p)_\Gg) \stackrel{\cong}{\leftarrow}
\hyper^*(\Hh,\complex^\times(p)_\Hh)$$
where the second isomorphism is induced by the map $\Gg \stackrel{M}{\to} \Hh$.
\end{lemma}

We will postpone the proof of this result to a forthcoming paper\cite{LupercioUribe3}.
This \u{C}ech description will be the one that will allow us understand the
relationship between the $B$-field and the discrete torsion and also gives us
an inside view of what the hypercohomology calculates.

We conclude this section by observing that with the definition of {\it gerbe
with connection} over an orbifold given in our previous work\cite{LupercioUribe1} 
we can easily prove the following

\begin{proposition}
For $\Gg$ a Leray description of a smooth \'{e}tale groupoid, a gerbe with
connection is a 2-cocycle of the complex $\breve{C}(\Gg,\complex^\times(3)_\Gg)$, that
is, a triple $(h,A,B)$ with $B \in \breve{C}(\Gg_0, \Aa^2_{\Gg,\complex})$,
$A \in \breve{C}(\Gg_1, \Aa^1_{\Gg,\complex})$ and
$h \in \breve{C}(\Gg_2, \complex^\times_\Gg)$ that satisfies $\delta B = dA$, 
$\delta A = d \log h$ and $\delta h = 1$. Two such gerbes with connection
are isomorphic if they lie in the same cohomology class, hence they are
classified by $\hyper^3(\Gg, \integer(3)_D^\infty)$.
\end{proposition}

\section{$B$-field and Discrete Torsion}

\subsection{Manifolds}
A $B$-field over a manifold $M$ (see Hitchin\cite{Hitchin})
is a choice of gerbe with connection which in terms
of a Leray cover $\{U_\alpha\}$ of $M$ is described by a collection of 2-forms $B_\alpha$
over each $U_\alpha$, 1-forms $A_{\alpha\beta}$ over the double intersections $U_{\alpha\beta}
:=U_\alpha \cap U_\beta$ and $\complex^\times$-valued functions $h_{\alpha\beta\gamma}$ over
triple overlaps $U_{\alpha\beta\gamma}$ satisfying
\begin{eqnarray*}
B_\alpha - B_\beta & = & dA_{\alpha \beta}\\
A_{\alpha \beta} + A_{\beta \gamma} -  A_{\alpha \gamma} & = & d \log h_{\alpha\beta\gamma}\\
h_{\alpha \beta \gamma}^{-1}h_{\alpha \beta \nu} h_{\alpha \gamma \nu }^{-1}
h_{\beta \gamma \nu} & = & 1.
\end{eqnarray*}
If we consider the description as a groupoid of $M$ given by the Leray cover $\{U_\alpha\}$:
$$\Mm_0 := \bigsqcup_\alpha U_\alpha \ \ \ \ \ \ \ \hbox{and} \ \ \ \ \ \ \
\Mm_1 := \bigsqcup_{\alpha\beta} U_{\alpha\beta}$$
and we collect the information of these functions as sections of the sheaves defined
in the previous section, in other words $B \in \breve{C}(\Mm_0, \Aa^2_{\Mm,\complex})$,
$A \in \breve{C}(\Mm_1, \Aa^1_{\Mm,\complex})$ and
$h \in \breve{C}(\Mm_2, \complex^\times_\Mm)$, it's easy to see that those equations become:
\begin{eqnarray*}
\delta B &=& dA\\
\delta A &=& d \log h\\
\delta h &=& 1
\end{eqnarray*}
then the triple $(h,A,B)$ determines a cocycle in the \u{C}ech complex
$\breve{C}(\Mm, \complex^\times(3)_\Mm)$ (see theorem 5.3.11\cite{Hitchin}); hence,
by lemma \ref{cech=hyper} we obtain

\begin{theorem}
A choice of $B$-field in a manifold $M$ determines a cocycle $(h,A,B)$ of the complex
$\breve{C}(\Mm, \complex^\times(3)_\Mm)$ and vice versa. Moreover, the isomorphism classes
of choices of $B$-field are classified by the cohomology class of $(h,A,B)$ in
the hypercohomology group $\hyper^2(M, \complex^\times_M \stackrel{d \log}{\To}
\Aa^1_{M, \complex} \stackrel{d}{\To} \Aa^2_{M, \complex})$.
\end{theorem}

Then, in view of the isomorphism \ref{isodelignecohomology} we conclude

\begin{cor}
The choice of $B$-field over a manifold $M$ is classified by the third Deligne
cohomology group $\hyper^3(M, \integer(3)^\infty_D)$ of the manifold.
\end{cor}

Here we should point out that we\cite{LupercioUribe1} have generalized the picture described
by Hitchin\cite{Hitchin} to the case of orbifolds, and hence the previous statements
remain true for the orbifold case.

\subsection{Discrete torsion}

In what follows we will argue that over an orbifold the discrete torsion is just
another choice of gerbe with connection, as is the $B$-field.
In other words, the discrete torsion and the $B$-field are both extreme cases of
the same picture, namely gerbes with connection; while the $B$-field only
takes into account the differentiable
structure of the orbifold, the discrete torsion one only  considers
the extra structure added by the action of the groups.

In the case of a global quotient $M/G$ with $G$ a finite group acting via diffeomorphisms
over $M$, we can take one  \'{e}tale groupoid that models it\cite{LupercioUribe1}, i.e.
$\Xx_0:=M$ and $\Xx_1:= M \times G$ with the natural source and target maps: $s(z,g)=z$
and $t(z,g)=zg$; denoting by $\bar{G}:= * \times G \twoarrows *$ the groupoid associated
to $G$ we have a natural morphism
\begin{eqnarray}\label{inclusionofgroup}
\Xx \to \bar{G}.
\end{eqnarray}

As $\bar{G}$ is formed by a discrete set of points we have that
$$H^2(G, \complex^\times) \cong
\hyper^2(\bar{G}, \complex^\times_{\bar{G}} \stackrel{d \log}{\To}
\Aa^1_{\bar{G}, \complex} \stackrel{d}{\To} \Aa^2_{\bar{G}, \complex}),$$
as there is a natural monomorphism induced by (\ref{inclusionofgroup})
$$\hyper^2(\bar{G}, \complex^\times_{\bar{G}} \stackrel{d \log}{\To}
\Aa^1_{\bar{G},\complex} \stackrel{d}{\To} \Aa^2_{\bar{G},\complex}) \to
\hyper^2(\Xx, \complex^\times_{\Xx} \stackrel{d \log}{\To}
\Aa^1_{\Xx,\complex} \stackrel{d}{\To} \Aa^2_{\Xx,\complex})$$
we get a map
$$H^2(G, \complex^\times) \to \hyper^2(\Xx, \complex^\times_{\Xx} \stackrel{d \log}{\To}
\Aa^1_{\Xx, \complex} \stackrel{d}{\To} \Aa^2_{\Xx, \complex}).$$
\begin{theorem} For the orbifold $\Xx=[M/G]$ the homomorphism
$$H^2(G,\complex^\times) \to \hyper^2(\Xx, \complex^\times_{\Xx} \stackrel{d \log}{\To}
\Aa^1_{\Xx, \complex} \stackrel{d}{\To} \Aa^2_{\Xx, \complex})$$
is injective. Therefore the choice of discrete torsion is a subgroup
of the equivalence classes of gerbes with connection over the groupoid, which
is classified by the third Deligne cohomology of the orbifold, namely
$$(2 \in \sqrt{-1})^{-2}\cdot \hyper^3(\Xx, \integer(3)_D^\infty).$$
\end{theorem}

Let $c:G\times G \to \complex^\times$ be a 2-cocycle and $\bar{c}:M\times G\times G \to
\complex^\times$, $\bar{c}(x,g,h) :=c(g,h)$ its image under the morphism. If $(\bar{c},0,0)=0$
in $\hyper^2(\Xx, \complex^\times(3)_\Xx)$ then there exist a map
$f:M\times G \to \complex^\times$ such that $\delta f = \bar{c}$ and $df=0$. As
$$(\delta f)(x,g,h) = f(x,g)f(x,gh)^{-1}f(x,h) = \bar{c}(x,g,h)=c(g,h)$$
we get that the cocycle $c$ is also exact; 
take $\sigma: G \to \complex^\times$ with $\sigma(g):=f(x,g)$ for any
fixed $x$, then $\delta \sigma = c$.

\section{Gerbes with connection and the Inertia groupoid}

In this section we are going to construct the holonomy bundle of a gerbe with connection,
which in the case of a groupoid will be a flat line bundle over the inertia groupoid. To do this
we need to recall some definitions.

\subsection{Line bundles with connection}

From theorem 2.2.12 of Brylinski\cite{Brylinski} we know that the group of isomorphism classes
of line bundles with connection over a manifold $M$ is canonically isomorphic to its second Deligne 
cohomology, namely
$$(2\pi \sqrt{-1})^{-1} \cdot \hyper^2(M, \integer(2)_D^\infty) \cong \hyper^1(M, \complex^\times_M 
\stackrel{d \log}{\To} \Aa^1_{M,\complex}).$$
The same result can be extended to cover smooth \'{e}tale groupoids, let's explain the idea.
For $\Gg$ a Leray description of a groupoid $\Gg$, a line bundle with connection over it is a 
morphism of groupoids $\rho:\Gg \to \overline{\complex^\times}$ and a 1-form $A$ over $\Gg_0$ such that
$$s^*A -t^*A = d \log \rho_1.$$
But $\rho$ is a morphism of groupoids if and only if $\delta \rho_1 =1$; considering  $\rho_1$ also
as an element of $\breve{C}(\Gg_1, \complex^\times_{\Gg})$ (recall that $\rho_1 : \Gg_1 \to \complex ^\times$);
i.e.
$$(\delta \rho_1)(g_1,g_2) = \rho_1(g_2)\rho_1(g_1g_2)^{-1}\rho_1(g_1)=1.$$

\begin{proposition}
The line bundle with connection $(\rho,A)$ over $\Gg$ represents a 1-cocycle in
the complex $\breve{C}(\Gg,\complex^\times(2)_\Gg)$ and its isomorphism class
is classified by the respective element in $\hyper^1(\Gg, \complex^\times_\Gg \stackrel{d \log}{\to}
\Aa^1_\Gg)$.
\end{proposition}

\subsection{The inertia groupoid}

The inertia groupoid $\wedge \Gg$ is defined in the following way:
$$\wedge \Gg_0 = \{a \in \Gg_1 | s(a)=t(a)\}$$
$$\wedge \Gg_1 = \{(a,b) \in \Gg_2 | a \in \wedge \Gg_0 \}$$
with $\sr(a,b)=a$ and $\tr(a,b)=b^{-1}ab$. Here we consider the case in which this groupoid is smooth.
We know (see theorem 6.4.2\cite{LupercioUribe2}) that a gerbe over an \'{e}tale groupoid
$\theta: \Gg_2 \to \complex^\times$ with $\delta \theta =1$ determines a line bundle over the 
inertia groupoid $\rho : \wedge \Gg \to \overline{\complex^\times}$ with
$$\rho_1(a,b) = \frac{\theta(a,b)}{\theta(b,b^{-1}ab)}$$
(we recommend the reader to see our previous paper\cite{LupercioUribe2} to get acquainted
 with the terminology).

Now we want to extend the previous result to define a line bundle with connection over the
 inertia groupoid
from a gerbe with connection.
\begin{lemma} \label{lemmagerbe->bundle}
Let $\Gg$ be a Leray description of a smooth \'{e}tale groupoid and $(h,A,B)$ a 2-cocycle
of $\breve{C}(\Gg, \complex^\times(3)_\Gg)$ (a gerbe with connection). Then the pair $(\rho,\nabla)$
where$$\rho(a,b):=\frac{h(a,b)}{h(b,b^{-1}ab)} \ \ \ \ \ \hbox{and} \ \ \ \ \ \nabla:= A|_{\wedge \Gg_0}$$
is a 1-cocycle of  the induced complex $\breve{C}(\wedge \Gg, \complex^\times(2)_{\wedge \Gg})$.
\end{lemma}

We just need to prove that $\delta \nabla = d \log \rho$. We have that in 
$\breve{C}(\wedge \Gg, \complex^\times(2)_{\wedge \Gg})$:
$$(\delta \nabla)(a,b) = \nabla(\sr(a,b))\cdot b -\nabla(\tr(a,b))= \nabla(a)\cdot b
 - \nabla(b^{-1}ab)$$
and in $\breve{C}(\Gg, \complex^\times(3)_\Gg)$:
$$(\delta A)(a,b) = A(a)\cdot b -A(ab) + A(b) = d \log h(a,b)$$
$$(\delta A)(b, b^{-1}ab) = A(b) - A(ab) +A(b^{-1}ab) = d \log h(b,b^{-1}ab).$$

By definition $\nabla(a) = A(a)$ and $\nabla(b^{-1}ab) = A(b^{-1}ab)$, so we get
$$(\delta \nabla)(a,b) = (\delta A)(a,b) - (\delta A)(b, b^{-1}ab) = d \log \frac{h(a,b)}{h(b,b^{-1}ab)} =
d \log \rho(a,b).$$
$\square$

In the same way as before, 1-cocycles in $\breve{C}(\Gg, \complex^\times(3)_\Gg)$ induce
0-cocycles in $\breve{C}(\wedge \Gg, \complex^\times(2)_{\wedge \Gg})$, so we get that there
is a morphism
\begin{eqnarray} \label{holonomymorphism}
H^2\breve{C}(\Gg, \complex^\times(3)_\Gg) \To H^1\breve{C}(\wedge \Gg, \complex^\times(2)_{\wedge \Gg})
\end{eqnarray} 
that extends to a morphisms in hypercohomology:
\begin{theorem}
Let $\Gg$ be a smooth \'{e}tale groupoid, then there exists a holonomy homomorphism
   $$\xymatrix{ \label{holonomymorphism2}
   \hyper^2(\Gg, \complex^\times_\Gg \stackrel{d \log}{\to} \Aa^1_{\Gg, \complex} \stackrel{d}{\to} \Aa^2_{\Gg, \complex}) \ar[r] \ar[d]^\cong &
   \hyper^1(\wedge \Gg, \complex^\times_{\wedge \Gg} \stackrel{d \log}{\to} \Aa^1_{\wedge \Gg, \complex}) \ar[d]^\cong\\
   (2\pi\sqrt{-1})^{-2}\cdot \hyper^3(\Gg, \integer(3)_D^\infty) \ar[r]^{\hbox{holo}} &
(2\pi\sqrt{-1})^{-1}\cdot \hyper^2(\wedge \Gg, \integer(2)_D^\infty)
   }$$
which to every gerbe with connection assigns a line bundle with connection over the 
inertia groupoid $\wedge \Gg$. Moreover,
this line bundle is flat.
\end{theorem}
Let's assume $\Gg$ is the Leray description of such a groupoid, from lemma \ref{cech=hyper} 
$$H^2\breve{C}(\Gg, \complex^\times(3)_\Gg) \cong 
\hyper^2(\Gg, \complex^\times_\Gg \stackrel{d \log}{\to} \Aa^1_{\Gg, \complex} \stackrel{d}{\to} \Aa^2_{\Gg, \complex})$$
and in the same manner as proposition 1.3.4 of Brylinski\cite{Brylinski}, there is a canonical homomorphism
$$H^1\breve{C}(\wedge \Gg, \complex^\times(2)_{\wedge \Gg}) \to 
\hyper^1(\wedge \Gg, \complex^\times_{\wedge \Gg} \stackrel{d \log}{\to} \Aa^1_{\wedge \Gg, \complex})$$
which together with the morphism (\ref{holonomymorphism}) implies (\ref{holonomymorphism2}).

Using the notation of lemma \ref{lemmagerbe->bundle} we know that $\delta B = dA$, or in other words, 
$B(s(g))-B(t(g))=dA(g)$ for $g \in \Gg_1$. Then is clear that for $a \in \wedge \Gg_0$ $dA(a) =0$, hence
$d\nabla =0$. The induced line bundle with connection over the groupoid is flat. $\square$

In the case that $\Gg$ is an orbifold, these induced flat line bundles over the inertia groupoid are
precisely what Ruan\cite{Ruan} denoted by {\it inner local systems} (see the last section of our previous 
paper\cite{LupercioUribe2}), then we can conclude

\begin{proposition}
A gerbe with connection over an orbifold $\Gg$ determines an inner local system\cite{Ruan} over
the twisted sectors $\widetilde{\Sigma_1 \Gg}$.
\end{proposition} 

The homomorphism of theorem \ref{holonomymorphism2}  can be generalized to all the 
Deligne cohomology groups. So we get the following result which will be proved in a
forthcoming paper\cite{LupercioUribe3}

\begin{theorem}
Let $\Gg$ be an smooth \'{e}tale groupoid, then there exist a natural morphism
of complexes 
$$\breve{C}(\Gg, \complex^\times(p)_\Gg) \to \breve{C}(\wedge \Gg, 
\complex^\times(p-1)_{\wedge \Gg})$$
which induces for all $p>0$ a morphism of cohomologies
    $$\xymatrix{
\hyper^q(\Gg, \complex^\times(p)_\Gg) \ar[r] \ar[d]^\cong &
 \hyper^{q-1}(\wedge \Gg, \complex^\times(p-1)_{\wedge \Gg}) \ar[d]^\cong\\
(2 \pi \sqrt{-1})^{-p+1} \cdot \hyper^{q+1}(\Gg, \integer(p)^\infty_D) \ar[r] &
(2 \pi \sqrt{-1})^{-p+2} \cdot \hyper^{q}(\wedge \Gg, \integer(p-1)^\infty_D).}$$
\end{theorem}

Observe that these facts provide a generalization of the concept of inner local system for the twisted multisectors of Ruan. It is reasonable to predict that this will be relevant in the full theory of Gromov-Witten invariants on orbifolds.

\section*{Acknowledgments}

We would like to thank conversations with M. Crainic, I. Moerdijk and 
Y. Ruan regarding several aspects of this work.

The first author would like to thank Paula Lima-Filho for some conversations that motivated his interest in Deligne Cohomology. 

The second author would like to express its gratitude to the organizers of the Summer
School on Geometric and Topological Methods for Quantum Field Theory, and specially to
its sponsors the Centre International de Math\'{e}matiques Pures et Appliqu\'{e}es (CIMPA)
and the Universidad de los Andes, for the invitation to take part in the school where
some of the previous results were presented.

\end{document}